\documentclass[runningheads]{llncs}
\usepackage{graphicx}
\usepackage{xcolor}
\usepackage{makecell}
\usepackage{changepage}
\usepackage{pdflscape}
\usepackage{afterpage}

\begin{document}
\title{What rationales drive architectural decisions? An empirical inquiry}
\titlerunning{Architectural decision rationales}

\author{Klara Borowa\inst{1}\orcidID{0000-0002-7160-5950} \and
        Rafał Lewanczyk\inst{1}          \and
        Klaudia Stpiczyńska\inst{1}  \and
        Patryk Stradomski\inst{1} \and
        Andrzej Zalewski\inst{1}\orcidID{0000-0001-5254-4761}}

\authorrunning{K. Borowa et al.}

\institute{Warsaw University of Technology, Institute of Control and Computation Engineering, Warsaw, Poland \\
\email{klara.borowa@pw.edu.pl}\\}

\maketitle              
\begin{abstract}
Architectural decision-making is a crucial concern for researchers and practitioners alike. There is a rationale behind every architectural decision that motivates an architect to choose one architectural solution out of a set of options. This study aims to identify which categories of rationale most frequently impact architectural decisions and investigates why these are important to practitioners. Our research comprises two steps of empirical inquiry: a questionnaire (63 participants) and 13 interviews. As a result, we obtained a set of rationales that motivated architects' decisions in practice. Out of them, we extracted a list of software quality attributes that practitioners were the most concerned about. We found that, overall, architects prefer to choose solutions which are familiar to them or that guarantee fast software implementation. Mid-career architects (5 to 15 years of experience) are more open to new solutions than senior and junior practitioners. Additionally, we found that most practitioners are not concerned about the quality attributes of compatibility and portability due to modern software development practices, such as the prevalence of using specific standards and virtualisation/containerization.

\keywords{Software Architecture  \and Architectural decision-making \and Rationale \and Software Quality Attributes}
\end{abstract}
\section{Introduction}

\color{black}
Understanding software architecture as a set of architectural decisions (ADs) \cite{jansen2005software} draws our attention to the motivation underlying these decisions and - this way - the entire architecture. Design rationale, which is a component of ADs 
\cite{zimmermann2009managing}, consists of the knowledge and reasoning justifying design decisions\cite{Tang2006}.


The research on factors (including rationales) \cite{Razavian2019} that shape architectural decisions in practice is rather scarce and seems still far from being mature.
The most recent papers by Weinreich et al. \cite{Weinreich2015}, Miesbauer et al. \cite{Miesbauer2013} and Tang et al. \cite{Tang2006} that explore the motivations underlying practitioners' ADs are at least eight years old. 
\color{black} These works are continued in more recent studies that investigate what software quality attributes (QAs) are discussed when choosing architectural patterns \cite{Bi2018} and what technology features drive technology design decisions \cite{Soliman2015}. 

As the software development landscape changes rapidly, the general purpose of this study is to discover what rationales, and why, currently drive ADs in practice. Such results importantly extend our knowledge and understanding of architectural decision-making (ADM)
\color{black}
by allowing researchers to focus their efforts on improving ADM on the basis of current needs and practices of architects. Additionally, we put an emphasis on QAs since they are a rationale subset that has been of major interest for researchers \cite{bhat2020evolution} \cite{Bi2018} \cite{Bi2021}.
\color{black}

Such an aim is expressed by the following research questions:
\begin{itemize}
    \item RQ1: What rationales most frequently influence architectural decisions? 
    \item RQ2: Which software quality attributes are usually prioritised during architectural decision-making? 
    \item RQ3: Why do practitioners prioritise these rationales?
\end{itemize}

In order to investigate the above problems we performed a two-phase inquiry. Firstly, we gathered data through a questionnaire. We obtained answers from 63 practitioners. Then, we presented the questionnaire's results to 13 practitioners during interviews. 
As a result of the questionnaire, we created a list of rationales (including quality attributes as given in ISO 25010 \cite{iso25010:2011}) that practitioners of various experience levels (beginners, mid-career and experts) consider essential.   \color{black} As a result of the interviews, we found out that, depending on experience level, practitioners tend to prioritise different architectural options.
\color{black}

The rest of the paper has been organised as follows: section \ref{sec:related_work} presents related work, Section \ref{sec:method} contains details about our research process and Section \ref{sec:resuts} the study's results. We discuss our findings in Section \ref{sec:discussion}, present the threats to validity in Section \ref{sec:ThreatsToValidity} and conclude in Section \ref{sec:Conclusion}.

\section{Related Work}
\label{sec:related_work}

\color{black}
The notion that software architecture is a set of design decisions \cite{jansen2005software} has heavily impacted the field of software architecture \cite{bhat2020evolution}. To enable better decision-making, researchers have explored such areas as: human factors in ADM \cite{Razavian2019}, AD models \cite{zimmermann2009managing}, mining AK \cite{Bi2021}, curating AK \cite{bhat2019adex}, tools supporting decision-making \cite{Liu2019}, techniques that can aid designers in the decision-making process \cite{Razavian2016} \cite{Tang2021}
\color{black}
and ADM rationale \cite{Tang2006}.


Numerous aspects make ADM an extremely challenging process. The traditional decision-making process, which includes listing all possible alternatives and their attributes, is impractical for software design decisions \cite{Burge2008} because of the number of possible architectural solutions. Furthermore, practitioners can be overwhelmed by the time and effort required to find architectural information \cite{Dieu2022}. Additionally, an entirely rational design-making process is impossible as long as it depends on human beings, that are impacted by various human factors \cite{Razavian2019}.

While there exist general guidelines \cite{Tang2021} and various tools \cite{bhat2020evolution} for ADM, empirical research on ADM factors is scarce \cite{Razavian2019}. On the topic of the practitioners' rationale behind design decisions, several studies must be acknowledged. Firstly, the study of Tang et al. \cite{Tang2006}, reporting the results of a survey on practitioners' approach to architectural rationale. Researchers had practitioners choose the importance of generic rationales and optionally allowed participants to provide their own rationales. As a result, a list of 12 rationales indicated by practitioners was made. This study's results were later expanded by Miesbauer et al. \cite{Miesbauer2013} and  Weinreich et al. \cite{Weinreich2015} who performed interview-based studies through which the list was expanded to include 18 rationales in total. Soliman et al. \cite{Soliman2015} researched what technology features impacted technology design decisions. Bi et al. \cite{Bi2018} took a different approach and researched which ISO 25010 software quality attributes \cite{iso25010:2011} were most often discussed in the context of architectural patterns on the StackOverflow platform.

\color{black}
We found no recent empirical research focusing widely on ADM rationale more recent than eight years ago. As software technology evolves rapidly, the rationales could also change. 
\color{black}
Additionally, we found no studies on how rationales depend on architects' professional experience, which we believe could be relevant since junior and senior architects find different aspects of ADM challenging \cite{tofan2013difficulty}.

\section{Method}
\label{sec:method}
Our research comprises two phases: questionnaire and interviews. The purpose of the questionnaire was to gather a larger sample of data that would enable us to answer RQ1 and RQ2. The interviews let us delve deeper into the meaning and implications of the questionnaire's results (RQ3). Another reason for using two data-gathering methods was to achieve so-called 'methodological triangulation' \cite{Runeson2012}, which helps to strengthen the validity of our findings. The overview of the study process is presented in Figure \ref{fig:method}. The questionnaire questions, a summary of questionnaire results, the interview plan, and interview coding details are available online \cite{additional_material}.

\afterpage{
\begin{figure}
\label{fig:method}
\includegraphics[width=17cm]{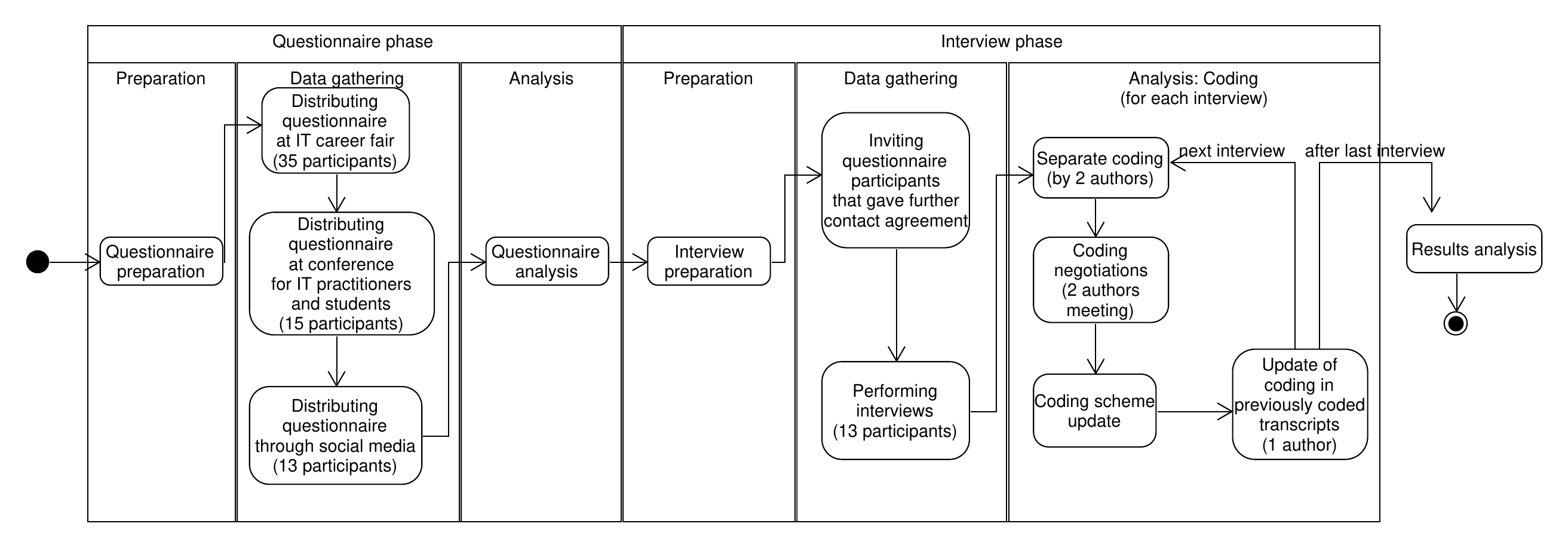}
\caption{Study phases}
\end{figure}
}

\subsection{Questionnaire: data-gathering}
\label{sec:questionnaire-data-gathering}
The questionnaire's \cite{additional_material} design was simplistic in order to avoid discouraging practitioners from taking part and to avoid biasing the results by suggesting any specific answers. The questionnaire was divided into four main sections:

\begin{enumerate}
    \item Participant data: age, gender, education, years of experience in software development, role in the company, company size, company domain.
    \item An open-ended question to provide a maximum of three most often used rationales for architectural decisions, according to the participant's personal experience.
     \item An open-ended question to provide a maximum of three most often used rationales for architectural decisions by the participant's colleagues. We asked this question to investigate if the participants believed that other practitioners have different priorities from them.
    \item An optional section containing the option to provide an email and give consent for further contact from the researchers.
\end{enumerate}

In order to obtain samples for the study, we distributed the questionnaires in three different locations: 
\begin{enumerate}
    \item During a 3-day long IT career fair at our faculty, where representatives of over 50 companies were present. We approached each stall and gave a physical copy of the questionnaire to the practitioners that were advertising their companies. We obtained 35 completed questionnaires at this event.
    \item During an IT conference for practitioners and students, where representatives from over 60 companies were present. We used the same strategy as the one during the career fair and obtained 15 additional completed questionnaires.
     \item We made the questionnaire available online and posted it on our personal social media accounts; this led to additional information from 13 participants.
\end{enumerate}
In total, we obtained data from 63 participants. A summary of the participants' demographic data is presented in Figure \ref{fig:participants}, and their employers' companies' domain and size in Figure \ref{fig:companies}.

\afterpage{
\begin{figure}
\label{fig:participants}
\includegraphics[width=\textwidth]{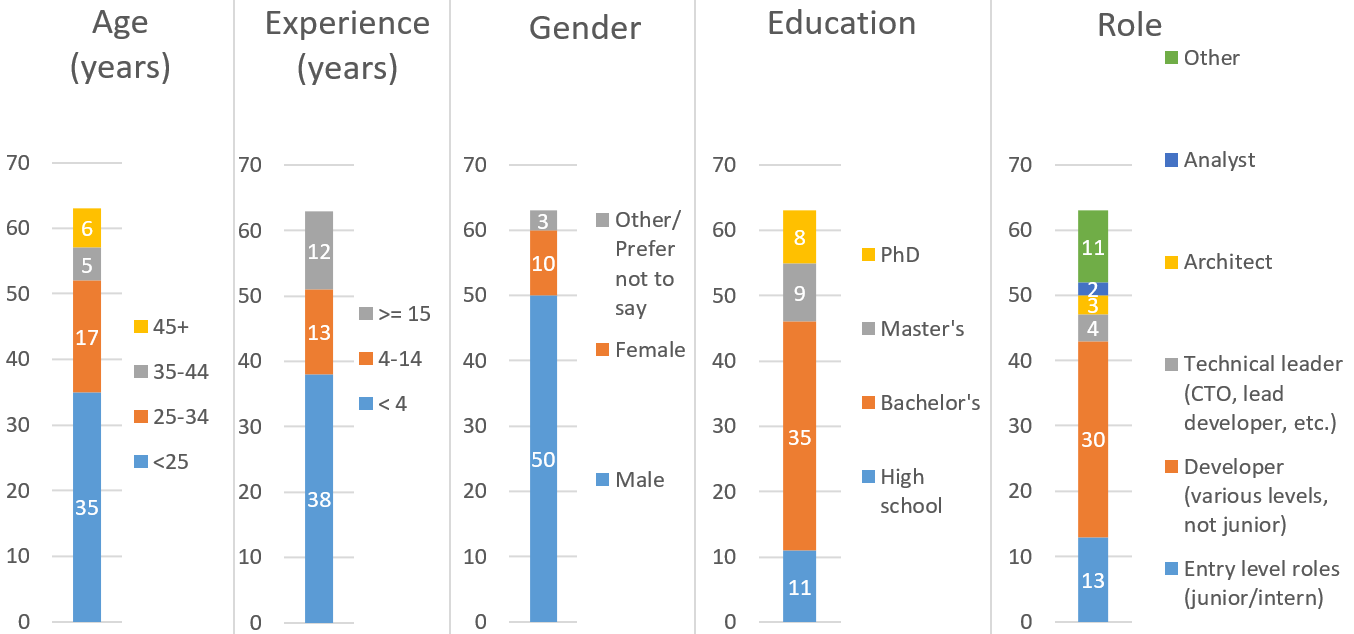}
\caption{Questionnaire participants} 
\end{figure}

\begin{figure}
\label{fig:companies}
\includegraphics[width=\textwidth]{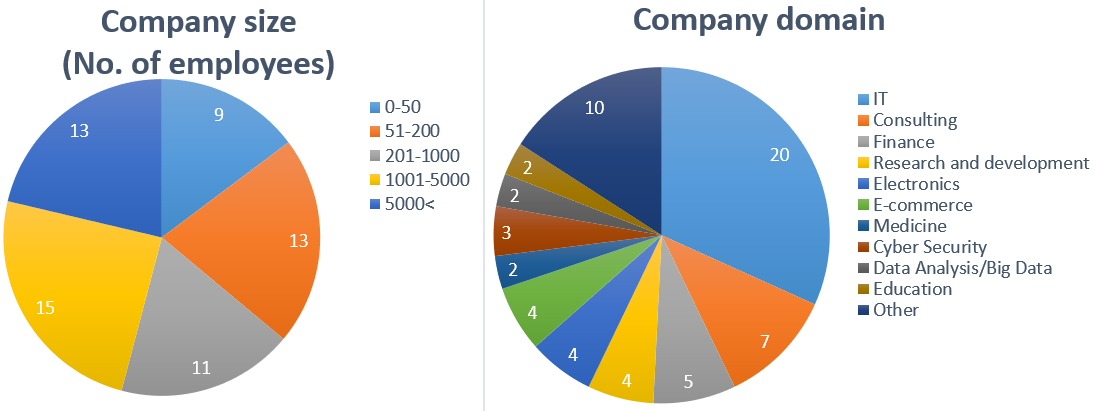}
\caption{Questionnaire participants companies} 
\end{figure}
}

\subsection{Questionnaire: analysis}
\label{sec:questionnaire-analysis}

To analyse the questionnaire, we performed the following actions:

\begin{enumerate}
    \item We divided the participants into the following groups: beginners (under five years of experience), mid-career (5 to 14 years of experience), and experienced (15 or more years of experience) practitioners.
    \item We extracted the answers about the participants' as well as their colleagues' rationales and analysed them separately. 
    \item For each of the six combinations of the above groups  (participants' experience level and their own/colleagues' rationales) separately, we classified the rationales (even if they were worded differently) into categories. When applicable, we used the ISO/IEC 25010 \cite{iso25010:2011} software quality attributes as the rationale categories. 
\color{black}
We grouped rationales into categories, since participants often used different words to explain the same factors influencing their decision-making. A rationale category groups rationales that are similar to such a degree that we found them almost indistinguishable.
\color{black}
For example, we categorised all of the following as ``Time/Deadlines'': ``time that we will waste on it; how much time there is to do it; time available to create the software; Number of hours required to write the functionality; time-consumption of making the solution; time-consuming; deadline to deliver the project; time available; time''. 
\color{black}
When rationales were only related \color{red}to each other\color{black}, like for example ``Documentation'' and ``Maintainability'', we did not categorise them together.
\color{black}
Table \ref{tab:questinaire-results} summarises the questionnaire analysis results.
\end{enumerate}

\begin{table}
\caption{Interview participants}\label{tab:interview-participants}
\begin{tabular}{|l|l|p{1cm}|p{1.5cm}|l|p{3cm}|p{2cm}|p{2cm}|}
\hline
No.	&	Gender	&	Age (years)	&	Experience (years)	&	Education	&	Role	&	Company size (employees)	&	Company domain
\\
\hline
1	&	Male	&	23	&	1	&	Bachelor's	&	Software Engineer	&	1001-5000	&	Infrastructure monitoring	\\
2	&	Male	&	22	&	1	&	Bachelor's	&	C++ Developer	&	51-200	&	Power Engineering	\\
3	&	Male	&	45	&	22	&	PhD	&	Company owner	&	0-50	&	IT, Data Science	\\
4	&	Male	&	23	&	1	&	Bachelor's	&	Pythin Backend Developer	&	51-200	&	Software House	\\
5	&	Male	&	22	&	1	&	High School	&	Junior Developer	&	1001-5000	&	E-commerce	\\
6	&	Male	&	23	&	3	&	Bachelor's	&	Junior Java Developer	&	over 5000	&	Consulting	\\
7	&	Male	&	24	&	4	&	Bachelor's	&	Software Engineer	&	51-200	&	Finance	\\
8	&	Male	&	31	&	5	&	Master's	&	Software Developer	&	1001-5000	&	Electronics	\\
9	&	Male	&	45	&	20	&	PhD	&	Architect	&	over 5000	&	Commerce	\\
10	&	Female	&	25	&	3	&	Master's	&	NLP Engineer	&	over 5000	&	R\&D 	\\
11	&	Male	&	41	&	20	&	PhD	&	CTO	&	201-1000	&	Finance	\\
12	&	Male	&	28	&	5	&	High School	&	Senior Testing Engineer	&	201-1000	&	Videogame development	\\
13	&	Male	&	32	&	6	&	Master's	&	Senior Software Engineering Manager	&	over 5000	&	FMCG	\\
\hline
\end{tabular}
\end{table}

\subsection{Interviews: data gathering}
\label{sec:interviews:data-gathering}
 Based on questionnaire data analysis, when creating the interview plan \cite{additional_material}, we focused on the following categories of observations:
\begin{enumerate}
    \item The rationales common for 20\% of the participants of each professional experience level. 
    \item Quality attributes of generally low interest to the architects, namely, attributes mentioned by fewer than 5\% of all the participants.
    \item Cases in which answers varied among architects of different experience levels. For example, some rationales were over the 20\% cutoff score in one group but not in all of them. 
\end{enumerate}
We presented the results from the questionnaire in which the above cases occurred to the interviewees. Then, we asked them about the reasons behind the observed level of importance of these rationales for specific architects' experience groups.

\color{black}
All 13 interviewees were recruited from the questionnaire participants. We invited to a follow-up interview all participants that consented to a follow-up interview in the questionnaire. 
\color{black}
Table \ref{tab:interview-participants} presents the overview of the interviewees' characteristics.

\afterpage{
\begin{table}
\caption{Codes}\label{tab:codes}
\begin{tabular}{|l|p{7cm}|p{2cm}|p{2cm}|}
\hline
Code	&	Description	&	Number of occurrences	&	Number of interviews where code occurred	\\
\hline
EX	&	Perspective/performed tasks change with the developer's experience	&	58	&	13	\\
CLNT	&	Recognising client's needs, focusing on the client's benefit.	&	31	&	12	\\
EASY	&	Participant mentions how important ease of use for development/maintenance is in the project	&	28	&	13	\\
FUT	&	Thinking about what effects the choice will have for the project	&	29	&	12	\\
D	&	Focusing on the deadline/ how much time something will take	&	23	&	9	\\
FAM	&	Choosing something based on one's familiarity with it	&	25	&	9	\\
IMP	&	The rationale was omitted because it is 'obviously' important	&	18	&	9	\\
EMP	&	Thinking how the choice will impact other people	&	15	&	10	\\
CR	&	Focusing on personal growth	&	15	&	8	\\
OUTDATED	&	The rationale does not require much thought because it is handled by newer technology	&	13	&	8	\\
NEG	&	The participant disagrees with other practitioners' opinions (from the questionnaire)	&	11	&	7	\\
EDU	&	Described behaviour is an effect of education	&	10	&	7	\\
RARE	&	The participant considers something as niche or unimportant	&	9	&	5	\\

\hline
\end{tabular}
\end{table}
}

\subsection{Interviews: analysis}
\label{sec:interviews:data-analysis}
The interview recordings have been transcribed. Then we coded the transcripts by following the subsequent steps:
\begin{enumerate}
    \item Two separate authors coded the same transcript using the descriptive coding method \cite{saldana2021coding}. This means that segments of the transcripts, which  contained a relevant piece of information, were labelled with a code that described its type of content.
    \color{black}
    We started with an empty list of codes, to avoid biasing the results towards our own ideas, and allowed the codes to emerge during the coding process.
    \color{black}
    \item Both coding authors met to negotiate their coding \cite{Garrison2006} --- they made changes to the coding until reaching a unanimous consensus. 
    \item An updated list of codes was created as a result of the coding meeting.
    \item One of the authors re-coded previously coded transcripts with new codes if they emerged during the current analysis step.
    \item The above steps were repeated for each interview transcript.
\end{enumerate}
Codes are summarised in Table \ref{tab:codes}. After coding all transcripts, we analysed and discussed the coded segments to draw conclusions.

\section{Results}
\label{sec:resuts}
Table \ref{tab:questinaire-results} presents the questionnaire results. As explained in Section \ref{sec:interviews:data-gathering}, we consider the rationale as important to a given group of architects if it was indicated but at least 20\% of them. Additionally, we focused on software quality attributes that were mentioned by less than 5\% of the participants and the variation in rationale prioritisation in different groups of participants.

\afterpage{
\pagestyle{plain}
\begin{landscape}
\begin{table}
\caption{Questionnaire results. ISO/IEC 25010 quality attributes are marked by a \textbf{bold} font.}\label{tab:questinaire-results}
\begin{tabular}{|l|l|l|l|l|l|l|l|l|l|}

\hline
 & &  \multicolumn{2}{|l|}{Sum} & \multicolumn{2}{|l|}{Beginners} & \multicolumn{2}{|l|}{Mid-career} & \multicolumn{2}{|l|}{Experienced}\\
\hline
No. & Rationale category	&	Participants	&	Colleagues	&	Participants	&	Colleagues &	Participants	&	Colleagues 	&	Participants	&	Colleagues\\
\hline

1	&	Ease of use for development	&	23	&	11	&	16	&	7	&	2	&	0	&	5	&	4	\\
2	&	\textbf{Maintainability	}&	15	&	2	&	12	&	1	&	2	&	1	&	1	&	0	\\
3	&	\textbf{Performance}	&	14	&	6	&	13	&	6	&	0	&	0	&	1	&	0	\\
4	&	Prior knowledge/experience	&	14	&	14	&	11	&	9	&	1	&	2	&	2	&	3	\\
5	&	Time/deadline	&	12	&	8	&	10	&	6	&	1	&	0	&	1	&	2	\\
6	&	\textbf{Reliability}	&	10	&	4	&	6	&	3	&	2	&	1	&	2	&	0	\\
7	&	Development Project Environment 	&	9	&	2	&	4	&	1	&	3	&	1	&	2	&	0	\\
8	&	Cost	&	8	&	9	&	5	&	7	&	1	&	0	&	2	&	2	\\
9	&	Popularity	&	8	&	8	&	7	&	5	&	0	&	1	&	1	&	2	\\
10	&	Scalability	&	7	&	3	&	4	&	3	&	2	&	0	&	1	&	0	\\
11	&	Business/customer requirements	&	7	&	5	&	4	&	4	&	1	&	0	&	2	&	1	\\
12	&	Documentation	&	6	&	4	&	6	&	4	&	0	&	0	&	0	&	0	\\
13	&	\textbf{Usability}	&	5	&	0	&	3	&	0	&	2	&	0	&	0	&	0	\\
14	&	\textbf{Security}	&	5	&	2	&	3	&	2	&	2	&	0	&	0	&	0	\\
15	&	Aesthetics/UX	&	5	&	2	&	1	&	1	&	2	&	0	&	2	&	1	\\
16	&	Fit with existing systems/project	&	5	&	7	&	4	&	4	&	0	&	1	&	1	&	2	\\
17	&	Decision-making methodology	&	5	&	4	&	0	&	0	&	2	&	1	&	3	&	3	\\
18	&	Testability (simplicity of writing tests)	&	4	&	0	&	3	&	0	&	0	&	0	&	1	&	0	\\
19	&	Level of complexity of the problem/system	&	4	&	1	&	4	&	1	&	0	&	0	&	0	&	0	\\
20	&	Expertise of more experienced colleagues	&	4	&	1	&	4	&	1	&	0	&	0	&	0	&	0	\\
21	&	\textbf{Functional Suitability}	&	3	&	1	&	2	&	1	&	0	&	0	&	1	&	0	\\
22	&	Availability of packages	&	3	&	0	&	1	&	0	&	0	&	0	&	2	&	0	\\
23	&	Team members' preferences	&	3	&	4	&	2	&	4	&	0	&	0	&	1	&	0	\\
24	&	\textbf{Portability}	&	2	&	2	&	2	&	1	&	0	&	0	&	0	&	1	\\
25	&	System life expectancy	&	2	&	0	&	0	&	0	&	0	&	0	&	2	&	0	\\
26	&	I want to add new skill to my resume	&	2	&	1	&	2	&	1	&	0	&	0	&	0	&	0	\\
27	&	\textbf{Compatibility}	&	1	&	2	&	0	&	0	&	0	&	0	&	1	&	2	\\
28	&	Return on Investment (ROI)	&	1	&	1	&	1	&	1	&	0	&	0	&	0	&	0	\\
29	&	Market expectations	&	1	&	0	&	1	&	0	&	0	&	0	&	0	&	0	\\
30	&	Available human resources/money	&	0	&	4	&	0	&	2	&	0	&	2	&	0	&	0	\\
31	&	Bus factor &	0	&	1	&	0	&	1	&	0	&	0	&	0	&	0	\\
32	&	``It works so I should use it''	&	0	&	1	&	0	&	1	&	0	&	0	&	0	&	0	\\
33	&	My colleagues have the same rationales as me	&	0	&	19	&	0	&	13	&	0	&	4	&	0	&	2	\\
\hline
\end{tabular}
\end{table}
\end{landscape}
}
\subsection{RQ1 \& RQ2: Most frequent rationales and prioritised software quality attributes}
The rationales that most frequently occurred in the questionnaires (over 20\% of participants) were:
\begin{enumerate}
    \item \textbf{``Ease of use for development''} was the dominant rationale for almost all groups of participants. Over 40\% of the beginner and expert groups believed that it was important. However, this was not the case for mid-career practitioners, where only 15\% mentioned this rationale.
    \item The quality attribute of \textbf{``Maintainability''} was the second most often indicated rationale, which was mentioned by 24\% of the participants. This was due to the beginners' insistence that this rationale is important (30\% of them), though it was not similarly prioritised by mid-career practitioners (15\%) and experts (8\%).
    \item Both the quality attributes of \textbf{``Performance''} and \textbf{``Prior knowledge/experience''}  were mentioned by the same number of practitioners overall (22\%). 
\textbf{``Performance''}, similarly to \textbf{``Maintainability''}, was important to beginners (33\%) but not to mid-career (0\%) and to expert practitioners (8\%).
\textbf{``Prior knowledge/experience''} of the solution, in the same way as \textbf{``Ease of use for development''}, was prioritised by both beginners and experts (over 20\% in both groups) but not by mid-career practitioners (only 8\%).
\end{enumerate}

Rationales that were overall mentioned by less than 20\% of the participants but were important for a particular group of practitioners (over 20\% of that group):
\begin{enumerate}
    \item \textbf{``Time/deadline''} is a rationale that was mentioned by 26\% of beginners but less often by mid-career and expert practitioners (8\% in both groups).
    \item \textbf{``Development Project Environment''}, which refers to various aspects of management and organisation of development project (e.g. company standards, client specifics) or current possibilities (available technologies), was important to mid-career practitioners (23\%) but less so to beginners (10\%) and experts (16\%).
    \item A \textbf{``decision-making methodology''} was by experts (25\%) but only a few mid-career practitioners (8\%) and no beginners.
\end{enumerate}

Three software quality attributes were mentioned by less than 5\% of the participants: \textbf{Compatibility} (1 participant), \textbf{Portability }(2 participants) and \textbf{Functional Stability} (3 participants). 

Finally, when asked about their colleagues' rationales, most participants wrote unprompted in their questionnaires that \textbf{their colleagues are motivated by the same rationales as they are themselves }(30\%). These were not cases of copying the same answers from one question to another but literally writing a statement about one's colleagues. This answer dominated the beginner (33\%) and mid-career (31\%) groups but occurred less frequently in the expert group (17\%).

\subsection{RQ3: Rationales' origins}
By analysing the interviews, we found a key set of rationales' origins.
\color{black}
Some rationales and rationale origins may slightly overlap (e.g. ``Time/deadlines'' rationale and ``fear of deadlines'' rationale origin). This was the case when participants listed both a rationale in the questionnaire and a rationale's origin in the interviews. 
\color{black}
The rationale's origins include (number of code occurrences overall/number of interviews where code occurred):
\begin{enumerate}
    \item \textbf{Practitioner's experience(58/13)}: The primary origin of the practitioners' rationales were their previous experiences. Beginners had limited experience, and to avoid the risk of not performing their duties efficiently, they preferred the solutions which they had used previously - because of that, ``Ease of use for development'' and ``Prior knowledge'' turned out to be the prevailing rationale for them. As one of the participants stated: ``(...) [junior developers] are such fresh people, it is certainly much more convenient. Because, well, since it's easy to learn something [how to use a solution], it's easy to reach the right level quite quickly.'' \par
    
    Experts with significant experience also prioritised these rationales but for different reasons -- they already had knowledge that they were confident in, so they did not feel the need to try new solutions and leave their comfort zone, e.g. ``Maybe more experienced people who worked a long time with a certain technology change it less often than people who are just entering the IT market (...), but feel comfortable with certain technologies and have been comfortable working with them for many years.''\par
    
    The exception to this effect were mid-career practitioners who were most likely to possess the knowledge and willingness to discover new  solutions. As a participant said: ``Maybe the moderately experienced people are neither those very experienced people who have been working in a particular technology for a longer period of time, but those who change it more often and maybe they see that it is not that difficult, they are used to changing technologies.''\par
    
    The practitioner's experience influence was also crucial for choosing the ``Time/deadline'' and ``decision-making methodology'' rationales. Beginners feared the possible consequences of missing a deadline more than other practitioners. Hence, they indicated the ``Time/deadline'' rationale more frequently than more experienced architects, e.g. ``People with more experience are more assertive when it comes to deadlines and are able to say `no' when they know that it is simply impossible to do something in a certain time, and those with less experience may also not be so sure that this is the moment that it is worth saying 'no' and not doing something, they are afraid of the deadline.''\par
    
    However, using a ``decision-making methodology'' as their rationale's foundation was only possible to experienced practitioners due to their greater knowledge, e.g. `` (...) we [the architects] are just getting used to such methodologies, acquiring them, so we will only use them after some time.''
    
    \item \textbf{Client focus (31/12)}: Various rationales originated from the endeavour to meet the client's needs. Practitioners often prioritised ``Ease of use for development'' and ``Time/deadline'' rationales because they strived to deliver new functionalities to the client as soon as possible, e.g. ``(...) recently there has been a lot of emphasis on time to market and deadlines for implementing individual functionalities, which are usually short.''\par
    
    Similarly, the ``Development Project Environment'' had to be considered to satisfy the client's needs. Even if two projects appeared to be the same, the environment often made a difference in its development. As one participant stated: ``Otherwise, seemingly the project sounds the same, but in practice, the client often wants something completely different than the previous one.''\par
    
    Additionally, ``Performance'' was seen generally as a key software quality attribute from the client's perspective, since weak software performance (software freezes, long waiting times, etc.) was seen as very problematic to the clients, e.g. ``(...) usually the performance of the system is related to the comfort of use, so it seems to me that this is also the reason why performance is an important criterion.''\par
    
    \item \textbf{Making one's life ``easy'' (28/13)}: Generally, practitioners choose solutions that they believed would make their work as effortless as possible. This was not only related to the ``Ease of use for development'' rationale but also ``Prior knowledge'' (the source of information about what is ``easy'') and ``Maintainability'' (minimisation of future work). As one participant stated: ``Some developers are lazy, which means that solutions that are easier to maintain often scale easier and perhaps require less work or less mental effort to add a new feature or to fix a bug.''\par
    
    \item \textbf{Thinking of the project's future (29/12)}: In general, practitioners were aware of the software life-cycle and knew that ``Maintainability'' could impact the amount of effort they would have to put into maintaining the system in the future. However, ``Ease of use for development'' was also a rationale impacted by this factor. Practitioners believed that if it is easy to use a given solution, it will also be easier to find, hire and train new employees that would work on the project in the future, e.g. ``(...)the ease of training new employees to work, whenever the software is easier to develop and is based on popular technology or the code is transparent, it is easier to introduce someone new here.''
    
    \item \textbf{Fear of deadlines (23/9)}: The fear of missing a deadline had a major impact on beginner practitioners. This was not the case for mid-career and expert practitioners since they already had experiences with missed deadlines in their careers and had the capacity to imagine how such a situation could be handled. For example: ``I think it's because the more experienced ones, I also know that this is how managers and programmers work, as well as project managers, that they know that this deadline is set with some reserve.''
    
    \item \textbf{Familiarity with a particular solution(25/9):} Prior experience with a particular solution was the main source of architectural knowledge. Since it is rarely possible to explore all the possible alternatives, prior experiences are the primary source of information, e.g. ``Architecture, all engineering, in general, is based on experience, and experience means things that we brokne in previous designs, in previous products. And on this experience, which looks so negative, but is nevertheless building our knowledge, we base what we create in the future.''.
    
    \item \textbf{``Obviousness'' (18/9)}: In the case of the ``Functional Stability'' quality attribute, some practitioners expressed the opinion that the importance of this rationale is simply obvious, and as such, there was no need to mention it in the questionnaire, e.g. ``It's [Functional Stability] also so mundane and part of such day-to-day work that maybe we don't tie it to the architecture.''.
    
    \item\textbf{Empathy (15/10)}: ``Ease of use for development'' and ``Maintainability'' were often prioritised because of the practitioners' awareness that their colleagues will have to maintain and further expand a system in the future, e.g. ``It should be done in such a way that I would not hurt myself or that it would not be painful for my colleagues to maintain. I see in this perhaps some form of empathy.''.
    
    \item \textbf{Personal growth (15/8)}: Mid-career practitioners did not prioritise ``Ease of use for development'' and ``Prior knowledge'' rationales, as beginners and experts did. Our participants pointed out that mid-career practitioners are in a specific professional situation where they can already feel confident in their basic knowledge (unlike beginners) but strive to learn about new solutions to further develop their careers (unlike experts). As one participant stated: ``(...) resume driven development, i.e. we choose those technologies that will look nice in the CV, or that will make us learn something.''.
    
    \item \textbf{New technology handles the problem (11/7)}: In the case of the ``Compatibility'', and ``Portability' quality attributes, practitioners believed that new technologies already solved most problems related to these rationales. In the case of ``Compatibility'', currently, existing standards are widely used, and compatibility problems are rare. As a participant stated: ``(...) because everything is somehow compatible with each other, only a matter of certain calling some services(...)''. 
    
    Similarly, the widespread use of virtualisation and containerisation solved most problems with ``Portability'', as a participant stated: ``(...) because practically everything can be uploaded, containerized''.
    
    \item \textbf{Practitioner's education(10/7)}: ``Performance'' was stated to be a rationale prioritised by beginner practitioners that recently finished their degrees in a field related to Software Engineering. This was due to the focus on the use of optimal data structures and algorithms during their studies, e.g. ``(...) during studies and in earlier educational programming, a lot of emphasis was placed on making these solutions work quickly. I even had one subject where we were judged on how many minutes it took to run a program, so it stuck in my head a bit.''.
    
    \item \textbf{Perception of the quality attribute as unimportant(9/5)}: Some participants stated that in the case of the projects that they worked on, ``Compatibility'' and ``Portability'' quality attributes were not important. For example, the project was targeted to work on a very specific platform, as the participant stated: ``(...)projects are created, for specific hardware or for specific platforms, not multi-platform solutions.''
\end{enumerate}


\section{Discussion}
\label{sec:discussion}
\color{black}

Two top rationales that were not quality attributes were ``Ease of use for development'' and ``Prior knowledge / experience''. This result is similar to the findings of Miesbauer et al. \cite{Miesbauer2013} and 
\color{black}
Weinreich et al. \cite{Weinreich2015} who found that the most influential rationale was ``Personal experience / Preferences''. 
\color{black}
This implies that the current trend of researching human factors in ADM \cite{Razavian2019} \cite{bhat2020evolution} is appropriate for further understanding and improving ADM. To be more specific, it seems that practitioners prioritise minimising their own and their colleagues' workload, both in the short and the long term. This fits with the principle of ``Simplicity -- the art of maximising the amount of work not done''  \cite{Beck2001Principles} from Agile software development. However, if done inappropriately, this can lead to consequences such as incurring architectural technical debt \cite{kruchten2019managing}.

The quality attributes of ``Maintainability'' and ``Performance'' were perceived as the most important out of the set of ISO 25010 software quality attributes \cite{iso25010:2011}. This matches the findings of Bi et al. \cite{Bi2018} who  found these to be the most often discussed quality attributes in the context of architectural patterns. We further explain this phenomenon since we found that beginner practitioners emphasise these rationales more than experts. In the case of ``Maintainability'', it seems that they wanted to avoid their own future workload, which may be perceived as an intimidating perspective. In the case of ``Performance'', beginners followed the knowledge acquired during their formal education and the emphasis of scholars on algorithmic efficiency.

Additionally, we found that practitioners in general do not put an emphasis on the quality attributes of ``Portability'' and ``Compatibility''. Modern technologies deliver solutions that well-address both these issues. In the case of ``Portability'', there are many efficient tools that resolve such problems: virtualisation, containerisation or frameworks for building multi-platform applications. Furthermore, in some fields (like developing console video games), the hardware on which the software will be run can be accurately predicted. Challenges with ``Compatibility'' have been overcome mostly through the standardisation of the technologies used by practitioners; for example, in the case of web applications, a REST API between the front-end and back-end layers is a predictable solution that most would choose by default.

Finally, we discovered that depending on experience level, practitioners have a significantly different mindset when it comes to ADM.  Beginners are greatly influenced by a fear of the unknown: they fear that it would be too hard to develop the software, or to maintain it later, to learn new solutions during the projects, and the consequences of unmet deadlines. Experts experience less fear of deadlines but put an emphasis on ease of development to make their colleagues' work easier and feel comfortable with their current practices. They were also the only group to use any decision-making methodologies, which they found natural if they gained enough knowledge. Lastly, mid-career practitioners are the most open to learning about new solutions and attempting not to use ones that are not considered ``easy'', to create bespoke solutions that would fit their clients the best.


\section{Threats to validity}
\label{sec:ThreatsToValidity}
        In this Section we describe three main kinds of threats to validity \cite{Runeson2012}:\newline   	
\textbf{Construct Validity}
To find the participants' rationales for architectural decisions, the possible methods of enquiry are either methods based on self-reporting or observation of the participants' work. We have chosen self-reporting methods (questionnaires and interviews) since that enabled us to obtain data from a greater number of practitioners. However, it is still possible that the participants' actual rationale may differ from those that they reported. For example, they may be impacted by cognitive biases \cite{zalewski2017cognitive} that they are not aware of. \newline
\textbf{Internal Validity}
To maximise the internal validity of our findings, the coding of the transcripts was always done independently by two authors. Then, both discussed the coding until they unanimously agreed on all codes. This was done to minimise the impact of the researcher's bias on the findings. 
\color{black}
However, it is possible that factors that we did not consider could play a role in practitioners' approach to decision-making, such as their company's size or domain.
\color{black}\newline
\textbf{External Validity}
We used convenience sampling since it is an extreme challenge to obtain a random generalisable sample of software practitioners. However, we strived to overcome this by providing data source triangulation \cite{Runeson2012}: we searched for participants from three different sources (two in-person events and one on social media). This resulted in a varied group of participants. 
\color{black}
Though, worth noting is that the sample may be biased towards less experienced practitioners, due to the majority of participants having less than 4 years of professional experience.
Additionally, since our results partially match results from previous studies \cite{Weinreich2015} \cite{Miesbauer2013} \cite{Bi2018}, it seems that our sample was big enough to give us outcomes also noticeable to other researchers.
\color{black}


\section{Conclusion}
\label{sec:Conclusion}
In this study, we performed a mixed methods two-step empirical inquiry into the practitioners' rationale behind their architectural decisions. The three main contributions of this study are as follows:
\begin{enumerate}
    \item A list of the most impactful rationales that influence practitioners' architectural decision-making;
    \item An exploration of these rationales' origin;
    \item The finding of how a practitioner's experience has a significant impact on how they make architectural decisions.
\end{enumerate}

Future research could employ different research techniques to further confirm or disconfirm our findings. A survey on a random generalisable sample would be beneficial, as well as observational studies on practitioners that would explore their decision-making in real-time.
In accordance to our findings, since experience level seems to be a major factor shaping who architects make their decisions researchers should take it into account during future research on ADM. 

Practitioners could benefit from our study by understanding better the way they and their colleagues develop software architectures. The observation on the influence of experience on ADM should also be reflected in shaping a team's structure, e.g. it would be prudent to focus on having mid-career (between 5 and 14 years of experience) practitioners in their teams when working on an innovative project.

\bibliographystyle{splncs04}
\bibliography{references}
%




\end{document}